\documentclass[conference]{IEEEtran}
\IEEEoverridecommandlockouts
\usepackage[export]{adjustbox}
\usepackage{caption}
\usepackage{array}
\usepackage{textcomp}
\usepackage{multirow}
\usepackage{tabularx,booktabs} 
\usepackage{verbatim}
\usepackage{makecell}
\usepackage{url}
\usepackage{relsize} 
\usepackage{color}
\usepackage{float}
\usepackage{enumitem}
\usepackage{bbm}
\usepackage{tikz}
\usepackage{bm}
\usepackage{stfloats}
\usepackage{hyperref}  
\usepackage[top=0.75in,bottom=1.02in,left=0.62in,right=0.62in]{geometry}
\usepackage{diagbox} 
\usepackage{cite}
\usepackage{amsmath,amssymb,amsfonts}
\DeclareMathOperator*{\argmax}{arg\,max}
\usepackage{algorithmic}
\usepackage{graphicx}
\usepackage{textcomp}
\usepackage{xcolor}
\def\BibTeX{{\rm B\kern-.05em{\sc i\kern-.025em b}\kern-.08em
    T\kern-.1667em\lower.7ex\hbox{E}\kern-.125emX}}
\usepackage{balance}
\begin{document}

\title{Energy-Efficient UAV-assisted LoRa Gateways: A Multi-Agent Optimization Approach}

\author{
  Abdullahi Isa Ahmed\textsuperscript{1}, 
  Jamal Bentahar\textsuperscript{2,3}, 
  El Mehdi Amhoud\textsuperscript{1} \\
  \textsuperscript{1}College of Computing, Mohammed VI Polytechnic University (UM6P), Benguerir, Morocco. \\
  \textsuperscript{2}Department of Computer Science, Khalifa University, Abu Dhabi, UAE. \\
  \textsuperscript{3}Concordia Institute for Information Systems Engineering, Concordia University, Montreal, Canada. \\
  Emails: {\{abdullahi.isaahmed, elmehdi.amhoud\}@um6p.ma, jamal.bentahar@ku.ac.ae}
}
\maketitle

\begin{tikzpicture}[remember picture,overlay]
\node[anchor=north,yshift=-20pt, text=black] at (current page.north) {\parbox{\dimexpr\textwidth-\fboxsep-\fboxrule\relax}{
\centering\footnotesize 
This paper has been accepted for publication in the 2026 IEEE 103rd Vehicular Technology Conference (VTC2026-Spring) \textcopyright 2026 IEEE.\\
Please cite it as: A.I. Ahmed, J. Bentahar, and E. M. Amhoud, “Energy-Efficient UAV-assisted LoRa Gateways: A Multi-Agent Optimization Approach,” in 2026 IEEE 103rd Vehicular Technology Conference (VTC2026-Spring), pp. 1–6, 2026.}};
\end{tikzpicture}

\begin{abstract}
As next-generation Internet of Things (NG-IoT) networks continue to grow, the number of connected devices is rapidly increasing, along with their energy demands, creating challenges for resource management and sustainability. Energy-efficient communication, particularly for power-limited IoT devices, is therefore a key research focus. In this paper, we study Long Range (LoRa) networks supported by multiple unmanned aerial vehicles (UAVs) in an uplink data collection scenario. Our objective is to maximize system energy efficiency by jointly optimizing transmission power, spreading factor, bandwidth, and user association. To address this challenging problem, we first model it as a partially observable stochastic game (POSG) to account for dynamic channel conditions, end device mobility, and partial observability at each UAV. We then propose a two-stage solution: a channel-aware matching algorithm for end device-UAV association and a cooperative multi-agent reinforcement learning (MARL) based multi-agent proximal policy optimization (MAPPO) framework for resource allocation under centralized training with decentralized execution (CTDE). Simulation results show that our proposed approach significantly outperforms conventional off-policy and on-policy MARL algorithms.
\end{abstract}


\vspace{-5pt}
\section{Introduction} \label{Section_1}

As 5G networks mature and 6G systems emerge, next-generation Internet of Things (NG-IoT) technologies are revolutionizing global connectivity by enabling massive machine-type communications (mMTC) across healthcare, smart cities, and autonomous systems \cite{jouhari2023survey, bazzi2502upper}. Furthermore, it is projected that the number of connected IoT devices will reach about 125 billion by 2030 \cite{fakhruldeen2025enhancing}. As these connected devices continue to proliferate, energy consumption has become a critical bottleneck for battery-powered devices, impacting not only network sustainability but also global climate objectives, including the United Nations Sustainable Development Goal 7 \cite{orazi2018first}.

Consequently, the deployment of low-power wide area networks (LPWANs), particularly long-range (LoRa) technology, has emerged as a promise solution for low-power, long-distance, and cost-effective communication in IoT applications. However, existing terrestrial LoRa networks depend on fixed ground-based gateways, which struggle with non-line-of-sight (NLoS) propagation. While deploying additional terrestrial LoRa gateways is affordable, it does not necessarily resolve NLoS issues. Conversely, satellite-based IoT solutions, such as the FOSSA system\footnote{FOSSA systems is a low-Earth orbit (LEO) satellite network providing global IoT connectivity for remote areas. More details at: https://fossa.systems/}, aim at connecting IoT devices to non-terrestrial networks. However, this approach significantly increases transmission power requirements and introduces higher latency, making it impractical for many energy-constrained IoT applications.

Beyond infrastructure deployment, effective resource allocation is essential for optimizing LoRa network performance. Existing studies largely rely on alternative optimization techniques, where complex problems are decomposed into sub-problems and solved iteratively. Although these methods can be effective in certain settings, they often fail to adapt to varying IoT environmental dynamics, resulting in suboptimal resource utilization. For example, the work in \cite{xiong2023flyinglora} applies an alternative optimization method for a single flying LoRa gateway, but lacks adaptability in dynamic environments. On the other hand, reinforcement learning (RL)-based methods, such as those proposed in \cite{yu2020multi, aihara2019q}, adopt the Q-learning technique for resource allocation. However, their reliance on static Q-tables makes them impractical for complex and rapidly changing IoT scenarios. To enhance system energy efficiency, the study in \cite{10279198} employs a deep RL proximal policy optimization (PPO) framework, but it is limited to a single gateway and only optimizes spreading factor (SF), and transmission power (TP), which restricts both scalability and flexibility.

In this paper, we investigate the optimization of system energy efficiency (EE) in an unmanned aerial vehicle (UAV)-assisted multiple LoRa gateways deployment under dynamic environments and ground-to-air (G2A) propagation. Specifically, we jointly optimize SF, TP, bandwidth (BW), and end-device (ED) association under partial observability. To address this problem, we model the system as a partially observable stochastic game (POSG), where multiple decision-makers interact within a dynamic environment, and each agent must act based solely on its own local, imperfect observations rather than possessing global knowledge of the system state. Based on this framework, we propose a two-stage solution: a channel-aware matching scheme for ED-UAV association and a multi-agent proximal policy optimization (MAPPO) framework for resource allocation. To the best of our knowledge, this is the first multi-agent reinforcement learning (MARL)-based approach for joint optimization in multi-agent systems. Our main contributions are:
\begin{itemize}
\item We formulate the joint resource allocation and association problem for EE maximization in UAV-assisted LoRa networks, accounting for ground LoRa ED mobility and realistic G2A channel propagation.
\item We model the problem as a POSG under partial observability and we propose a two-stage solution combining channel-aware ED-UAV matching and a MAPPO-based centralized training with decentralized execution (CTDE) framework.
\item Our simulations result show that our approach requires fewer environment steps to achieve convergence compared to other state-of-the-art MARL algorithms. Additionally, during execution phase, our approach improved the system EE compared to both on-policy and off-policy RL.
\end{itemize}
The rest of this paper is organized as follows. In Section~\ref{Section_2}, the system model and optimization problem are introduced. In Section~\ref{Section_3}, we describe our proposed two-stage solution. The simulation setup and results are discussed in Section~\ref{Section_4}, and finally, we conclude the paper in Section~\ref{Section_5}.

\section{System Model} \label{Section_2} 
\begin{figure}[t]
  \centering 
\includegraphics[width=1.0\linewidth]{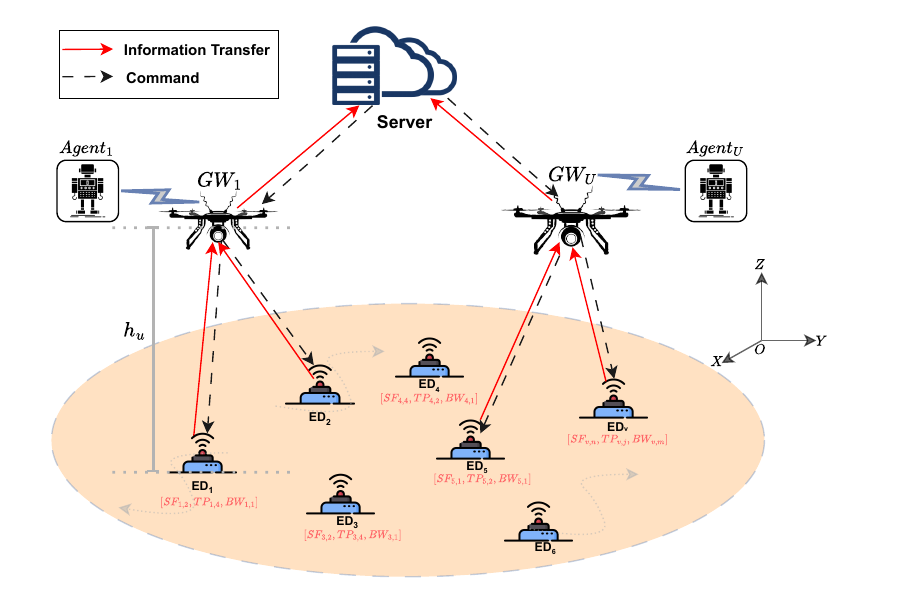} 
  \caption{The studied system model.}
  \label{fig_system_model}
  \vspace{-14pt}
\end{figure}

We consider an uplink data collection scenario in a LoRa network consisting of $\mathcal{V}$ LoRa EDs, $\mathcal{U}$ flying gateways, and a single network server. In this setup, we suppose that the UAVs are equipped with LoRa gateways deployed over a square target area $S$. Specifically, each LoRa gateway has a limited communication range $R_{comm}$ and can simultaneously connect to multiple EDs within its association quota $\Lambda_{max}$. Additionally, each gateway collects and decodes packets from all associated EDs, then relays these packets to the network server. The sets of EDs and gateways are denoted by $\mathcal{V} = \{1, \ldots, V\}$ and $\mathcal{U} = \{1, \ldots, U\}$, respectively. The system model is illustrated in Fig.~\ref{fig_system_model}.

\subsection{Mobility Model}\label{subsec:mobility}
In this work, we consider a Gauss-Markov (GM) mobility model \cite{zhang2025novel} for ground LoRa EDs. Specifically, each ED $v \in \mathcal{V}$ is initially placed uniformly at random within the target area $S$ and assigned an initial velocity vector $\mathbf{v}_v[0] \in \mathbb{R}^2$. At each time $t \in \mathcal{T} =  \{1, \ldots, T\}$, the velocity is updated according to
\begin{equation}
\mathbf{v}_v[t] = \eth \, \mathbf{v}_v[t-1] + (1-\eth) \, \bar{\mathbf{v}}_v + \hat {\sigma} \sqrt{1-\eth^2} \, \mathbf{w}_t,
\label{eq:gm}
\end{equation}
where $\eth$ is the parameter of tuning that controls temporal correlation, $\bar{\mathbf{v}}_v$ is the asymptotic mean velocity of the ED $v$, and $\mathbf{w}_t \sim \mathcal{N}(\mathbf{0},\mathbf{I}_2)$ is independent and identically distributed as a standard bi-variate Gaussian noise. The randomness level is governed by $\hat {\sigma} \geq 0$. Furthermore, the ED position $\mathbf{p}_v[t]$ at time $t$ is then updated as
\begin{equation}
\mathbf{p}_v[t] = \mathbf{p}_v[t-1] + \mathbf{v}_v[t] \, \Delta t.
\end{equation}

To ensure physical consistency, the velocity magnitude is constrained to a maximum speed of $v_{\max}$. Additionally, when an ED reaches the boundary of $S$, the velocity component normal to the boundary is reversed to model reflection. With small probability at each step, $\bar{\mathbf{v}}_v$ is re-sampled uniformly from $[-v_{\max}, v_{\max}]^2$ to allow gradual long-term direction changes.

\subsection{Communication Model}\label{subsec:channel}
The ground-to-air (G2A) channel between UAV gateway $u$ and ED $v$ at time $t$ follows the probabilistic path-loss model, which differentiates between line-of-sight (LoS) and NLoS propagation \cite{saliah2024multi}. Specifically, the LoS probability depends on the elevation angle $\theta_{u,v}[t] \triangleq \tan^{-1}\left(\frac{h_u}{d_{u,v}[t]}\right)$ and is given by
\begin{equation}
\label{eq:plos}
P_{\text{LoS}}(\theta_{u,v}[t]) = \frac{1}{1 + \vartheta  \exp(-\lambda (\theta_{u,v}[t] - \vartheta ))},
\end{equation}
where $d_{u,v}[t] = \sqrt{(x_u - x_v[t])^2 + (y_u - y_v[t])^2}$ is the horizontal distance at time $t$, $h_u$ is the fixed UAV altitude, and $\vartheta$, $\lambda$ are environment-dependent parameters. Therefore, the NLoS probability is simply $P_{\text{NLoS}}(\theta_{u,v}[t]) = 1 - P_{\text{LoS}}(\theta_{u,v}[t])$.

Consequently, the average path loss at time $t$ is expressed as
\begin{equation}
\begin{split}
    \label{eq:a2g_pl}
L_{u,v}^{\text{G2A}}[t] = 20\log_{10}\left(\frac{4\pi f d_{u,v}[t]}{c}\right) + \eta_{\text{LoS}} P_{\text{LoS}}(\theta_{u,v}[t]) + \\ \eta_{\text{NLoS}} P_{\text{NLoS}}(\theta_{u,v}[t]),
\end{split}
\end{equation}
where $f$ is the carrier frequency, $c$ is the speed of light, $\eta_{\text{LoS}}$, and $\eta_{\text{NLoS}}$ are the additional attenuation factors for LoS and NLoS links, respectively.

Given the average path loss between the UAV and ED, their channel gain can be expressed as 
$G_{u,v}[t] = 10^{-L_{u,v}^{\text{G2A}}[t]/10}$.  Thus, the signal-to-noise ratio (SNR) between UAV $u$ and ED $v$ at time slot $t$ is given by $\rho_{u,v}[t] = \frac{P_{v}[t] \cdot G_{u,v}[t]}{\sigma^{2}}$, where $P_{v}[t]$ is the transmit power of ED $v$ at time $t$, and $\sigma^{2}$ denotes the noise power. In this work, we assume perfect orthogonality among different SFs used by LoRa EDs, meaning that interference occurs only from transmissions using the same SF. Under these assumptions, the signal-to-interference-plus-noise ratio (SINR) $\mho_{u,v}^{n}[t]$ experienced by UAV $u$ from ED $v$ when using the $n$-th SF at time slot $t$ is calculated as
\begin{equation}
    \mho_{u,v}^{n}[t] = \frac{\rho_{u,v}[t]}{\sum_{v' \in \mathcal{V} \setminus \{v\}} \psi_{v',n}[t] \cdot \rho_{a_{v'}[t],v'}[t] + 1},
    \label{sinr}
\end{equation}
where $\rho_{a_{v'}[t],v'}[t]$ is the SNR between ED $v'$ and its serving UAV $a_{v'}[t]$, and $\psi_{v',n}[t] \in \{0,1\}$ indicates whether ED $v'$ employs SF $n$.

The achievable data rate for the link between UAV $u$ and ED $v$ at time slot $t$ can be expressed as
\begin{equation}
\Re_{u,v}[t] = W_{v}[t] \cdot \log_2\left(1 + \mho_{u,v}^{n}[t]\right),
\label{eqt_shannon}
\end{equation}
where $W_{v}[t]$ is the allocated bandwidth.
\subsection{UAV Power Model}

In this work, we consider the hover power consumption $P^{\text{hover}}_{u}$ of UAV $u$ for a multi-rotor platform as modeled in \cite{gong2023modeling}
\begin{equation}
P^{\text{hover}}_{u} = \bar{n} \times W_{r}^{3/2} \times \left( \hat{\rho}^{-1/2} s A^{-1/2} C_{T}^{-3/2} \frac{\delta}{8} + \frac{(1+k)}{\sqrt{2 \hat{\rho} A}} \right),
\label{eqt_hover_power}
\end{equation}
where $\bar{n}$ is the number of rotors, $W_r$ is the per-rotor weight, $\hat{\rho}$ is air density, $s$ is the solidity ratio, $A$ is the rotor disc area, $C_T$ is the thrust coefficient, $\delta$ is the blade profile drag coefficient, and $k$ is the induced power factor. Unless otherwise stated, the parameter values used to evaluate the hover power follow those reported in \cite{gong2023modeling}. 

\subsection{Association and Resource Allocation} \label{signal_model1}
We consider binary association between UAV $u$ and ED $v$ at time slot $t$ is denoted by $a_{u,v}[t]  \in \{0, 1\}$ and is defined as
\begin{equation}
a_{u,v}[t]  = 
\begin{cases}
1, & \text{if ED } v \text{ is served by UAV } u \text{ at time } t, \\
0, & \text{otherwise}.
\end{cases}
\label{eq:association_binary}
\end{equation}
Note that the index of the UAV selected by ED $v$ at time $t$ can be expressed as $a_{v}[t]  = \sum_{u \in \mathcal{U}} a_{u,v}[t]  \cdot u$.

Beyond association, each ED must be allocated appropriate transmission parameters from discrete sets of spreading factors, transmission powers, and bandwidths. Specifically, the SF is selected from a vector $\mathbf{\Psi} = \{\psi_{1}, \dots, \psi_{N}\}$. For each ED $v$, the allocation of SF follows a binary association expressed as $\psi_{v,n}[t] $, $n \in \mathcal{N} = \{1, \dots, N\}$, where $\psi_{v,n}[t]  = 1$ if ED $v$ communicates at SF $\psi_{n}$ during time slot $t$, $\psi_{v,n}[t] = 0$ otherwise. The selected index is $\Psi_{v}[t]  = \sum_{n \in \mathcal{N}} \psi_{v,n}[t]  \cdot \psi_{n}$. Assuming each ED transmits using only one SF at any given time, we impose the constraint
\vspace{-5pt}
\begin{equation}
    \sum_{n=1}^{N} \psi_{v,n}[t]  \leq 1, \quad \forall v \in \mathcal{V}, t \in \mathcal{T}.
    \label{sf_binary}
\end{equation}

Moreover, the transmission power level is selected from a vector $\mathbf{P} = \{p_{1}, \dots, p_{J}\}$ in dBm. For each ED $v$, we define a binary allocation variable $p_{v,j}[t] $, $j \in \mathcal{J} = \{1, \dots, J\}$, where $p_{v,j}[t]  = 1$ if ED $v$ transmits with power level $p_{j}$ at time $t$, and $p_{v,j}[t]  = 0$ otherwise. The selected index is $P_{v}[t]  = \sum_{j \in \mathcal{J}} p_{v,j}[t]  \cdot p_{j}$. Each ED transmits using only one TP at each time step, imposing the constraint
\vspace{-5pt}
\begin{equation}
    \sum_{j=1}^{J} p_{v,j}[t]  \leq 1, \quad \forall v \in \mathcal{V}, t \in \mathcal{T}.
    \label{tp_binary}
\end{equation}

Furthermore, the communication bandwidth $W$ is selected from the set $\mathbf{W} = \{w_{1}, \dots, w_{M}\}$ in kHz. We assume that each deployed gateway operates with a specific LoRa bandwidth from this set. Hence, the binary variable $w_{v,m}[t] $, $m \in \mathcal{M} = \{1,\dots, M\}$, indicates the bandwidth selected by ED $v$ at time $t$, where $w_{v,m}[t]  = 1$ if ED $v$ transmits using bandwidth $w_{m}$, and $w_{v,m}[t]  = 0$ otherwise. The resulting selected bandwidth is $W_{v}[t]  = \sum_{m=1}^{M} w_{v,m}[t]  \cdot w_{m}$.

Consequently, we define finite sets for all possible BW allocations $\bar{\mathbf{W}}$, SF selections $\bar{\mathbf{\Psi}}$, TP allocations $\bar{\mathbf{P}}$, and ED-UAV association $\mathbf{a}$, which can be expressed as
\vspace{-7pt}
\begin{subequations}
\begin{align}
\bar{\mathbf{W}} = \{ W_{v}[t] \in \textbf{W} |\sum_{m=1}^{\mathcal{M}} w_{v,m}[t] \leq 1 , \quad \forall v \in \mathcal{V}\}, \label{w} \\
\bar{\mathbf{\Psi}} = \{ \Psi_{v}[t] \in {\bf \Psi} \mid \sum_{n=1}^{\mathcal{N}} \psi_{v,n}[t] \leq 1 , \quad \forall v \in \mathcal{V}\}, \label{sf} \\
\bar{\mathbf{P}} = \{ P_{v}[t] \in \textbf{P} |\sum_{j=1}^{\mathcal{J}} p_{v,j}[t] \leq 1 , \quad \forall v \in \mathcal{V}\},  \label{tp} \\
\mathbf{a} = \{ a_{v}[t] \in \mathcal{U} |\sum_{u=1}^{\mathcal{U}} a_{u,v}[t] \leq 1 , \quad \forall v \in \mathcal{V}\}. \label{nodes}
\end{align}
\end{subequations}
\begin{table}[t]
    \centering
    \caption{SNR $\rho_{\text{thr}}(\Psi_v, W_v)$ in dB \cite{9272315}}
    \label{table:snr_threshold}
    \small
    \begin{tabular}{|c|c|c|c|c|c|c|} \hline
        \diagbox{$\mathbf{W}$}{$\mathbf{\Psi}$} & 7 & 8 & 9 & 10 & 11 & 12 \\ \hline
        125 & -7.5 & -10 & -12.5 & -15 & -18 & -21 \\ \hline
        250 & -9 & -12 & -14.5 & -17 & -20 & -23 \\ \hline
        500 & -11 & -13.8 & -16.5 & -19 & -21.8 & -25 \\ \hline
    \end{tabular}
    \vspace{-10pt}
\end{table}
\subsection{Optimization Problem}
We model our system's total EE as
\begin{equation}
   EE_{\mathrm{sys}} = \sum_{t=1}^{T} \sum_{u=1}^{U} \left[ \; \frac{ \sum_{v=1}^{v} a_{u,v}[t]  \cdot  \Re_{u,v}[t] }{ \left( \sum_{v=1}^{v} a_{u,v}[t] \cdot  P_{v}[t]  \right) +P^{\text{hover}}_{u}} \; \right],
   \label{sys_ee}
\end{equation}
where $\sum_{v=1}^{v} a_{u,v}[t] \cdot  P_{v}[t]$ is the total uplink TP for all associated EDs. Therefore, to maximize system EE $EE_{\mathrm{sys}}$ while respective LoRa quality-of-service, we jointly optimize BW allocations $\bar{\mathbf{W}}$, SF allocations $\bar{\boldsymbol{\Psi}}$, TP selections $\bar{\mathbf{P}}$,  and ED-UAV associations $\mathbf{a}$. Our optimization problem is formulated as
\begin{subequations}\label{formulated_algo1}
\begin{align}
\max_{\bar{\mathbf{W}}, \bar{\boldsymbol{\Psi}}, \bar{\mathbf{P}}, \mathbf{a}} 
&  \quad \text{EE}_{\mathrm{sys}} \label{constr:a} \\
\text{s.t.} \quad 
& \;\sum_{u \in \mathcal{U}} a_{u,v}[t] \leq 1, \; \forall v \in \mathcal{V}, t \in \mathcal{T}, \label{constr:b} \\ 
& \;\sum_{v \in \mathcal{V}} a_{u,v}[t] \leq \Lambda_{\max}, \; \forall u \in \mathcal{U}, t \in \mathcal{T}, \label{constr:c} \\
& \;a_{u,v}[t] \in \{0, 1\}, \; \forall v \in \mathcal{V}, u \in \mathcal{U}, t \in \mathcal{T}, \label{constr:d} \\
& \;\psi_{v,n}[t] \in \{0, 1\}, \; \forall v \in \mathcal{V}, n \in \mathcal{N}, t \in \mathcal{T}, \label{constr:e} \\
& \;p_{v,j}[t] \in \{0, 1\}, \; \forall v \in \mathcal{V}, j \in \mathcal{J}, t \in \mathcal{T}, \label{constr:f}\\
& \;w_{v,m}[t] \in \{0, 1\},  \; \forall v \in \mathcal{V}, m \in \mathcal{M}, t \in \mathcal{T}, \label{constr:g} \\
& \;\rho_{u,v}[t] \geq  \rho_{\text{thr}}(\Psi_{v}[t], W_{v}[t]), \; \forall v \in \mathcal{V}, u \in \mathcal{U}, t \in \mathcal{T}, \label{constr:h} \\
& \; \eqref{sf_binary} \text{ and } \eqref{tp_binary}.
\end{align}
\end{subequations}
Constraint~\eqref{constr:b} ensures each ED associates with at most one UAV. Constraint~\eqref{constr:c} limits the number of EDs per UAV to the capacity $\Lambda_{\max}$. Constraints~\eqref{constr:d}--\eqref{constr:g} enforce binary decisions for ED-UAV association, SF, TP, and bandwidth selection, respectively. Constraint~\eqref{constr:h} guarantees the received SNR exceeds the minimum threshold for successful LoRa packet detection given the selected SF $\Psi_{v}[t]$ and the allocated BW $W_{v}[t]$, as specified in Table~\ref{table:snr_threshold} and given in \cite{9272315}.

\section{Proposed Solution} \label{Section_3}
The optimization problem in \eqref{formulated_algo1} is known to be NP-hard due to the combinatorial nature of binary association variables coupled with non-convex constraints, leading to an exponential growth in the solution space that conventional optimization methods cannot efficiently solve. To address this challenge, we propose a two-stage approach. First, we employ a channel-aware matching algorithm that dynamically assigns each ED to the UAV with the strongest channel gain, subject to communication range and capacity constraints. Second, given these associations, a MAPPO-based framework optimizes resource allocation for all associated EDs to maximize system-wide EE. The following subsections describe each stage in detail.

\subsection{ED-UAV Association Algorithm} \label{signal_model}
The ED-UAV association is managed through a matching algorithm executed at each time $t$. For each ED $v \in \mathcal{V}$, we compute the path gain $G_{u,v}[t]$ to all UAVs within range and select the UAV with maximum gain
\vspace{-5pt}
\begin{equation}
\begin{split}
a_v[t] = \argmax_{u \in \mathcal{U}} \Big\{ G_{u,v}[t] \;\Big|\; d_{u,v}[t] \leq R_{\text{comm}}, \\
\sum_{v' \in \mathcal{V}} a_{u,v'}[t] < \Lambda_{\max} \Big\},
\end{split}
\label{eq:association}
\end{equation}
where $d_{u,v}[t]$ is the distance between ED $v$ and UAV $u$. The binary association is set as $a_{u,v}[t] = 1$ if $u = a_v[t]$, and $a_{u,v}[t] = 0$ otherwise. Thus, the computational complexity is \(\mathcal{O}(U)\) per ED and \(\mathcal{O}(VU)\) for all \(V\) EDs.

\subsection{POSG Formulation}
In this work, we transform the resource allocation problem in \eqref{formulated_algo1} as a POSG \cite{albrecht2024multi} defined by the tuple $(\mathcal{U}, \mathcal{S}, \{\mathcal{A}_u\}_{u\in\mathcal{U}}, \{\mathcal{O}_u\}_{u\in\mathcal{U}}, \mathcal{T}, \{\mathcal{Z}_u\}_{u\in\mathcal{U}}, \{\mathcal{R}_u\}_{u\in\mathcal{U}}, \gamma, \mu_0)$, where each UAV acts as a decentralized agent under partial observability. Each component is defined as follows

\textbf{Agents $\mathcal{(U)}$:} The set of flying LoRa gateways $\mathcal{U}$.

\textbf{Action Space $\{\mathcal{A}_u\}$:} The agent takes an action $a_u[t] \in \mathcal{A}_u$ at time $t$, according to its local observation $o_u[t]$, which consists of three resource allocation parameters, which are the spreading factor $\psi[t]$, the transmission power $p[t]$, and the bandwidth $w[t]$. Thus, the action of UAV $u$ is expressed as $a_u[t] = \{\psi[t], p[t], w[t]\}$. The joint action at time $t$ is $\mathbf{a}[t] = \left( a_1[t], a_2[t], \dots, a_{U}[t] \right) \in \mathcal{A} = \prod_{u \in \mathcal{U}} \mathcal{A}_u$.

\textbf{Observation Space $\{\mathcal{O}_u\}$:} At each time $t$, agent $u$ observes only a local subset of the global state $\mathbf{s}[t] \in \mathcal{S}$, specifically the EDs currently associated with it within its communication range $R_{\text{comm}}$ and subject to its capacity constraint $\Lambda_{max}$. The partial observability arises from two factors: (i) each UAV only observes EDs associated with it, not all EDs in the system, and (ii) each UAV can serve at most $\Lambda_{max}$ EDs simultaneously. Local observation $o_u[t] \in \mathcal{O}_u$ is a matrix that contains, for the EDs associated with the highest $\Lambda_{max}$ and can be expressed as $o_u[t] = (x_v[t], y_v[t], d_{u,v}[t], G_{u,v}[t])_{\Lambda_{max} \times 4}$, where $d_{u,v}[t] = \|\mathbf{p}_v[t] - \mathbf{p}_u\|_{2}$ is the horizontal distance between the ED $v$ and UAV $u$, and $G_{u,v}[t]$ is the corresponding path gain computed using the G2A LoRa channel model. The joint observation\footnote{If UAV $u$ is associated with fewer than $\Lambda_{max}$ EDs, we introduced virtual EDs by padding with zeros to maintain consistent neural network input size.} is $\mathbf{o}[t] = (o_1[t], \dots, o_U[t])$.

\textbf{State Transition Function $\mathcal{(T)}$:} The transition function $\mathcal{T}(\mathbf{s}[t], \mathbf{a}[t], \mathbf{s}[t+1]) = \Pr(\mathbf{s}[t+1] \mid \mathbf{s}[t], \mathbf{a}[t])$ governs the evolution of the global state. That is, each ED $v$ updates its position stochastically using the GM mobility model in Eq.\eqref{eq:gm}, where the Gaussian noise $\mathbf{w}_t$ introduces randomness in ED trajectories. The ED-UAV associations $a_{u,v}[t]$ are updated using the channel-aware matching scheme in Eq.\eqref{eq:association}, ensuring optimal connectivity based on current channel conditions and UAV capacity constraints.

\textbf{Observation Function $\{\mathcal{Z}_u\}$:} The observation function $\mathcal{Z}_u: \mathcal{S} \times \mathcal{O}_u \rightarrow [0,1]$ specifies the probability of receiving observation $o_u[t]$ given the global state $\mathbf{s}[t]$, i.e., $\mathcal{Z}_u(\mathbf{s}[t], o_u[t]) = \Pr(o_u[t] \mid \mathbf{s}[t])$. The observation $o_u[t]$ contains information about EDs within communication range $R_{\text{comm}}$ and subject to capacity constraint $\Lambda_{max}$.

\textbf{Reward Function:} In this work, all agents share the same reward $\mathcal{R}_u(\mathbf{s}[t], \mathbf{a}[t]) = r[t], \;\forall u \in \mathcal{U}$, making the POSG fully cooperative. The instantaneous reward at time $t$ is designed to directly maximize system EE while promoting reliable communication and penalizing excessive power consumption. Therefore, we define our agent's reward function as
\begin{equation}
    r[t] = \omega_1 \cdot \text{EE}_{\text{sys}}[t] 
    + \omega_2 \cdot \Xi [t] 
    + \omega_3 \cdot \beta [t] 
    - \omega_4 \cdot P_{\text{total}}[t],
    \label{eq:reward}
\end{equation}
where $\text{EE}_{\text{sys}}[t]$ is the system EE given in Eq.~\eqref{sys_ee}, $\Xi[t] \in [0,1]$ is the communication success rate, $\beta[t]$ is a shaped term that rewards positive average SNR margin and heavily penalizes negative margins which address constraint~\eqref{constr:h}, $P_{\text{total}}[t]$ is the instantaneous total power consumption of the system, and $\omega_1,\omega_2,\omega_3,\omega_4$ are weighting coefficients. This shared reward ensures all UAVs collaboratively optimize the global objective despite relying only on their partial local observations $o_u[t]$.

Additionally, the discount factor $\gamma \in (0,1)$ balances immediate and future rewards, while the initial state distribution $\mu_0: \mathcal{S} \rightarrow [0,1]$ specifies the probability over initial states, where $\mathbf{s}_0 \sim \mu_0(\cdot)$ determines the starting ED positions and association configurations. Each UAV agent $u$ follows a policy $\pi_u: \mathcal{O}_u \times \mathcal{A}_u \rightarrow [0,1]$, parameterized by $\phi_u$, mapping its local observation to a probability distribution over actions, with the joint policy denoted as $\boldsymbol{\pi} = (\pi_1, \pi_2, \ldots, \pi_U)$. Hence, the objective of each agent is to maximize its expected cumulative discounted reward:
\begin{equation}
J_u(\pi_u) = \mathbb{E}_{\tau_u} \left[ \sum_{t=0}^{T} \gamma^t r[t] \right],
\end{equation}
where the expectation is taken over trajectories $\tau_u$ of agent $u$. In this fully cooperative setting, all agents aim to maximize the shared team objective $J(\boldsymbol{\pi}) = \sum_{u=1}^{U} J_u(\pi_u)$.

\subsection{MAPPO-based Resource Allocation} 
With ED-UAV associations established by the channel-aware matching algorithm in Section \ref{signal_model}, the next step is to optimize resource allocation across the associated ED-UAV pairs. We leverage MAPPO for this task, as it is well-suited for our POSG formulation, where each UAV observes only its associated EDs within $R_{\text{comm}}$ and $\Lambda_{max}$ constraints, rather than the complete global state.

Furthermore, MAPPO operates under the CTDE paradigm. Therefore, during our training phase, a centralized critic accesses the global state $\mathbf{s}[t]$, which includes all ED positions, channel conditions, and associations, to accurately evaluate joint actions $\mathbf{a}[t]$ based on the shared reward $r[t]$ in Eq.~\eqref{eq:reward}. Simultaneously, each UAV $u$ learns a decentralized policy $\pi_u$ mapping its local observation $o_u[t]$ to resource allocation actions $a_u[t] = \{\psi[t], p[t], w[t]\}$.

In addition, our agent's policy update uses a clipped objective function that compares the probability ratio between new policy $\pi_u$ and old policy $\pi_u^{old}$, restricting it to $[1-\epsilon, 1+\epsilon]$ to prevent destabilizing updates. Moreover, the advantage function $\hat{A}_u[t]$ guides improvement by measuring action quality relative to expected performance, promoting actions that increase system EE and communication reliability while managing power consumption. The critic minimizes the error between its predictions $V_\theta(\mathbf{s}[t])$ and empirical returns $V^{target}[t]$, with advantages computed using GAE to balance estimation bias and variance. Entropy regularization encourages exploration during training.

During execution, each UAV operates independently using only its policy $\pi_u(\cdot | o_u[t])$ and local observation $o_u[t]$, without inter-UAV communication or global state access. This decentralized approach eliminates coordination overhead and enables rapid response to local conditions while maximizing the shared team objective $J(\boldsymbol{\pi})$. Details on the MAPPO algorithm and its computational complexity are discussed in \cite{li2024collaborative}.
\begin{table}[h]
    \centering
    \caption{Simulation setup.}
    \vspace{-0.5em}
    \label{tab:my-table-parameter}
    \begin{adjustbox}{max width=0.98\columnwidth}
    \begin{tabular}{c|c|c|c}
        \toprule
        \textbf{Sym.} & \textbf{Value} & \textbf{Sym.} & \textbf{Value} \\
        \midrule
        $h_{u}$ & $90$ m   & $\textbf{P}$ & $[2, 5, 8, 11, 14]$ dBm \\
        $f$ & $868$ MHz & $\textbf{W}$ & $[125, 250, 500]$ KHz  \\
        $c$ & $3 \times 10^8$ m/s & $\mathbf{\Psi}$ & $[7, 8, 9, 10, 11, 12]$  \\
        $\vartheta$, $\lambda$ & $4.88$, $0.43$&  $\eta_{\text{LoS}}$, $\eta_{\text{NLoS}}$ & $0.1$ dB, $21$ dB  \\
        $\Delta t$, $\eth$ & $0.5$ s, 0.85 & $\hat {\sigma}$, $\sigma^{2}$  & 0.5, $-120$ dBm \\
        $\mathbf{\bar{v}}_v$ & $0.005$ & $v_{max}$ & $1.0$ m/s \\
        $\alpha$, $\gamma$, $\epsilon$ & $10^{-4}$, $0.99$, $0.2$ & Epochs & $15$  \\
        $T_{\max}$, $T$ & $2$M, $150$ & Batch size, $\tau$ & $16, 0.01$  \\
        $|\mathcal{D}|$, hidden$_{dim}$ & $32$, $128$ & Architecture & GRU  \\
        Optimizer & Adam  & Activation & ReLU \\
        Seed & $\{0, 42, 2021\}$  & $\omega_1,\omega_2,\omega_3,\omega_4$  & $4 \times 10^{-4}, 5.0, 1.0, -10^{-2}$  \\
        \bottomrule
    \end{tabular}
    \end{adjustbox}
\end{table}
\vspace{-10pt}
\section{Simulation Results} \label{Section_4}
\subsection{Simulation setup}
To evaluate the performance of our proposed approach, we consider an area of $1000\text{m} \times 1000\text{m}$, where ground LoRa EDs are initially distributed uniformly at random and move according to the Gauss-Markov mobility model. We assume the deployed UAVs hover at an altitude of $90\text{m}$. Specifically, the UAVs are placed evenly along the horizontal axis, each separated by equal spacing at the center of the deployment area to ensure balanced coverage. The simulation parameters are summarized in Table~\ref{tab:my-table-parameter}.

As benchmarks, we compare our approach with three widely used MARL algorithms: COMA \cite{foerster2018counterfactual}, VDN \cite{sunehag2017value}, and QMIX \cite{rashid2020monotonic}. COMA is an on-policy actor-critic method that employs a centralized critic with a counterfactual baseline to address the multi-agent credit assignment problem. VDN is an off-policy value decomposition approach that factorizes the joint action-value function into a sum of individual agent value functions, enabling decentralized training. QMIX, also an off-policy approach, extends value decomposition by enforcing a monotonic relationship between the global Q-function and the individual agent Q-functions, thereby ensuring consistency between centralized training and decentralized execution.
\begin{figure}[t]
  \centering 
\includegraphics[width=0.85\linewidth]{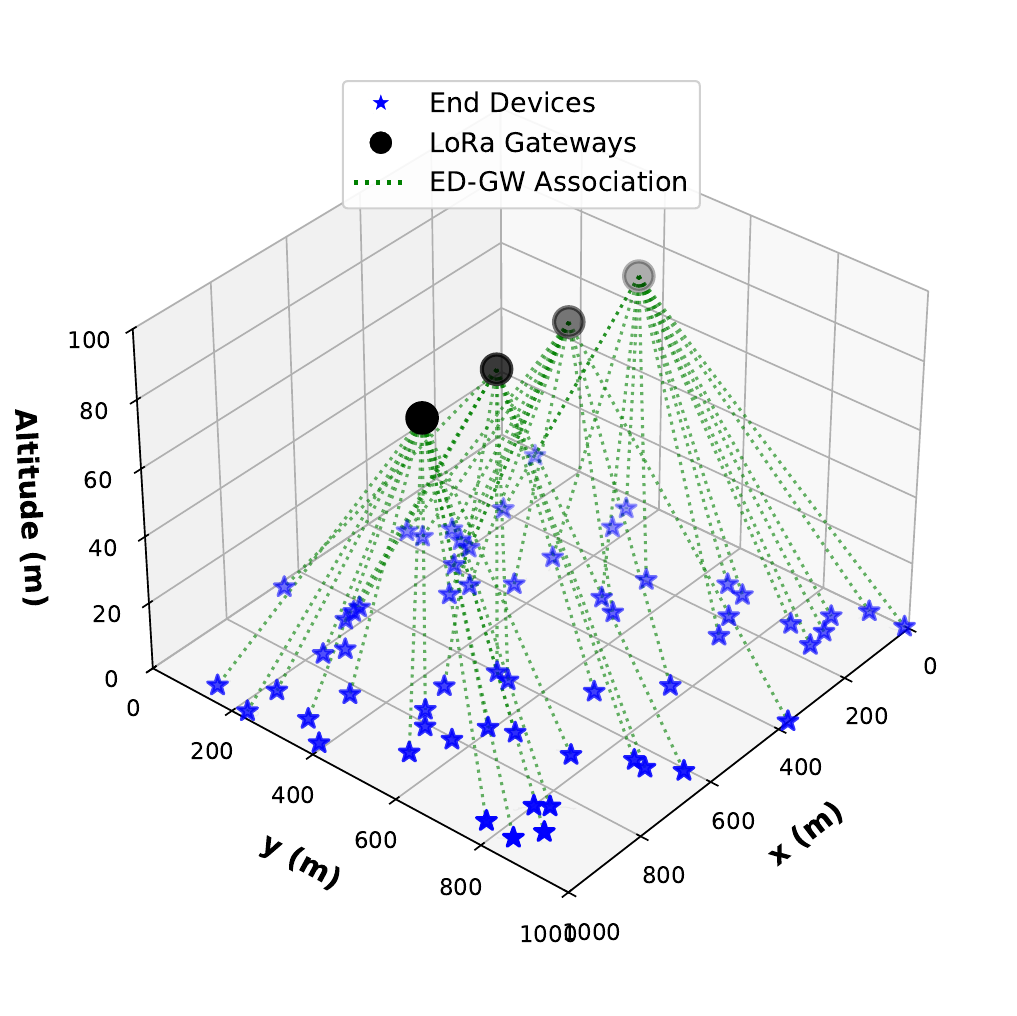} 
  \caption{Final ED-UAV association}
  \label{fig_ed_uav_association}
  \vspace{-14pt}
\end{figure}
\begin{figure*}[t]
\centering
\begin{tabular}{ccc}
    \vspace{-6pt}
    \includegraphics[scale=0.45]{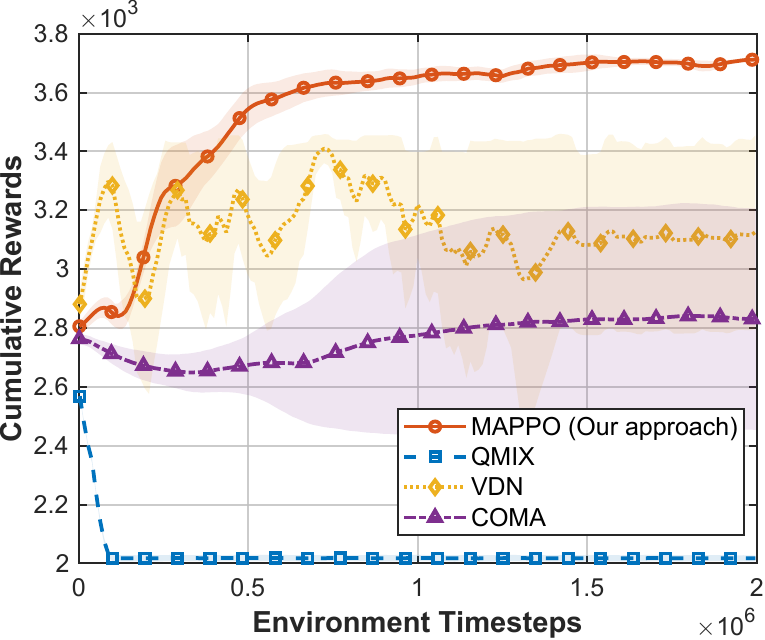} & 
    \includegraphics[scale=0.45]{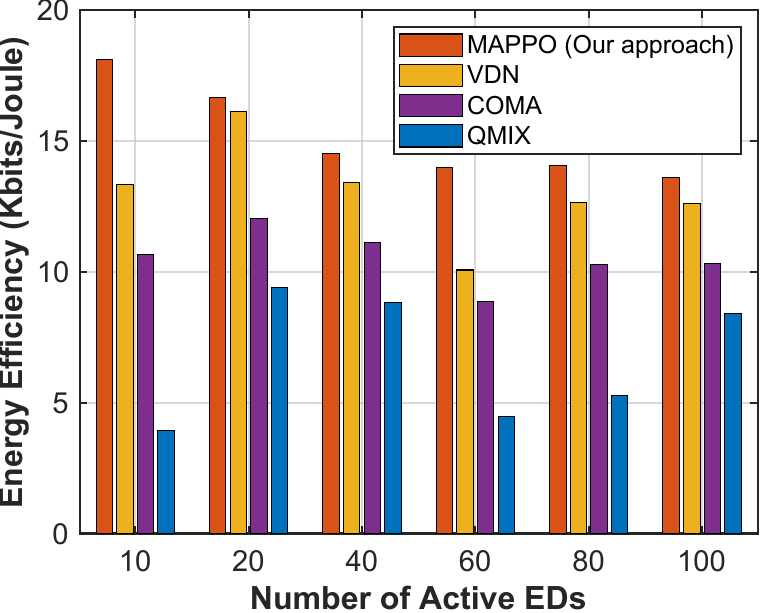} & 
    \includegraphics[scale=0.45]{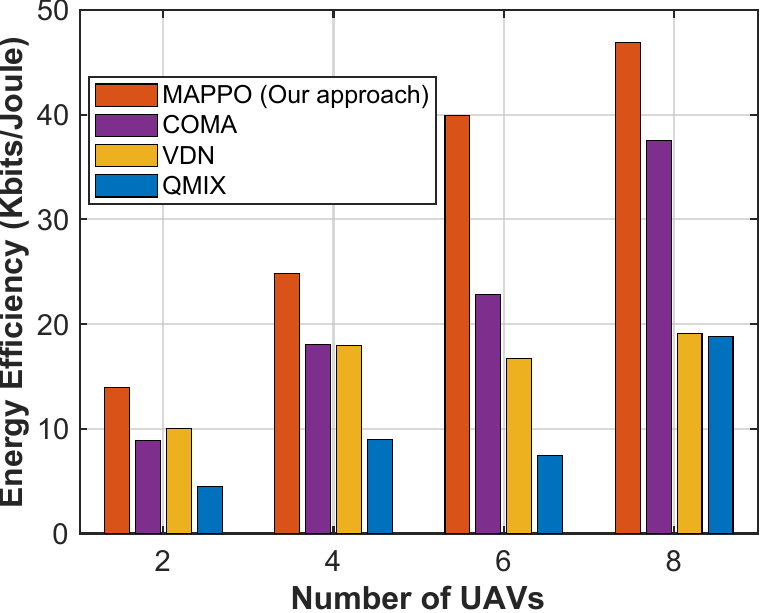} \\
    \small (a) & 
    \small (b) & 
    \small (c)
\end{tabular}
\caption{(a) Cumulative Rewards, (b) The performance of EE under different varying EDs, (c) EE comparison with varying gateways.}
\label{fig:training_optimization}
\end{figure*}

\subsection{Simulation Results}
Figure \ref{fig_ed_uav_association} presents the final association between the EDs and the UAV-assisted LoRa gateways obtained with our proposed channel-aware association scheme after convergence. Specifically, the scheme assigns EDs to gateways by maximizing the overall channel gain while respecting the quota and capacity constraints of each gateway. As shown in the figure, every ED is successfully associated with at least one gateway. In Fig. \ref{fig:training_optimization}(a), we plot the cumulative rewards over environment timesteps for all compared algorithms with 2 UAVs and 10 active EDs. As training progresses, our proposed MAPPO approach converges faster and obtains higher rewards as compared to other state-of-the-art MARL benchmarks. 

In Fig. \ref{fig:training_optimization}(b), we present the EE obtained during the execution phase using our trained agents with 2 UAVs and a varying number of active EDs. As expected, the figure presents the final average EE over three seeds, and it shows that our proposed MAPPO approach consistently achieves the highest performance across all numbers of active EDs. Compared to the second-best algorithm, VDN, our approach improves EE by 36.02\%, 3.29\%, 8.30\%, 39.01\%, 11.29\%, and 7.91\% for 10, 20, 40, 60, 80, and 100 active EDs, respectively. Our approach also outperforms QMIX by more than a factor of two in most cases. Additionally, EE exhibits a slight downward trend as the number of active EDs increases, which is expected because serving more active EDs increases the total system power consumption, causing power demand to grow higher. 

Finally, in Fig. \ref{fig:training_optimization}(c), we plot the EE against the varying number of deployed UAVs, ranging from 2 to 8, with 60 active EDs. As shown in the figure, most of the algorithms exhibit an upward trend in EE. This behavior is expected, since adding more UAVs increases coverage and improves link quality, which increases the achievable throughput more than the corresponding rise in power consumption. In addition, the improvement can also be attributed to the fact that EDs gain more flexibility to associate with UAVs that maximize the association utility function.

\section{Conclusion} \label{Section_5}
In this paper, we studied the problem of system energy efficiency maximization in UAV-assisted LoRa networks by jointly optimizing the spreading factor, transmission power, bandwidth, and UAV-ED association. To solve this problem, we proposed a two-stage solution: a channel-aware matching algorithm for ED-UAV association and a MAPPO-based framework for resource allocation under the CTDE scheme. Simulation results show that our proposed MAPPO approach significantly outperforms state-of-the-art MARL algorithms in terms of fast convergence and system energy efficiency. Future directions include integrating trajectory optimization for multiple UAV-mounted LoRa gateways to enable adaptive coverage and exploring hybrid deployments where LoRa operates in both the 868 MHz and 2.4 GHz bands.

\vspace{-3.3pt}
\section*{Acknowledgement}
This work was sponsored by the Junior Faculty Development program under the UM6P-EPFL Excellence in Africa Initiative.

\balance
\bibliographystyle{IEEEtran}
\bibliography{biblio}

@article{jouhari2023survey ,
  title={{A survey on scalable LoRaWAN for massive IoT: Recent advances, potentials, and challenges}},
  author={Jouhari, Mohammed and Saeed, Nasir and Alouini, Mohamed-Slim and Amhoud, El Mehdi},
  journal={IEEE Communications Surveys \& Tutorials},
  year={2023}
}

@article{fakhruldeen2025enhancing,
  title={{Enhancing smart home device identification in WiFi environments for futuristic smart networks-based IoT}},
  author={Fakhruldeen, Hassan Falah and others},
  journal={International Journal of Data Science and Analytics},
  volume={19},
  number={4},
  pages={645--658},
  year={2025},
  publisher={Springer}
}

@inproceedings{orazi2018first ,
  title={{A first step toward an IoT network dedicated to the sustainable development of a territory}},
  author={Orazi, Gilles and others},
  booktitle={Global Internet of Things Summit},
  year={2018}
}

@article{xiong2023flyinglora,			
title={{FlyingLoRa: Towards energy efficient data collection in UAV-assisted LoRa networks}},			
author={Xiong, Runqun and Liang, Chuan and Zhang, Huajun and Xu, Xiangyu and Luo, Junzhou},			
journal={Computer Networks},			
volume={220},			
pages={109511},			
year={2023},			
publisher={Elsevier}			
}

@inproceedings{yu2020multi ,
  title={{Multi-agent Q-learning algorithm for dynamic power and rate allocation in LoRa networks}},
  author={Yu, Yi and Mroueh, Lina and Li, Shuo and Terr{\'e}, Michel},
  booktitle={IEEE Inter. Symp. on Personal, Indoor and Mobile Radio Communications},
  year={2020}
}

@article{aihara2019q ,
  title={{Q-learning aided resource allocation and environment recognition in LoRaWAN with CSMA/CA}},
  author={Aihara, Naoki and Adachi, Koichi and Takyu, Osamu and Ohta, Mai and Fujii, Takeo},
  journal={IEEE Access},
  volume={7},
  pages={152126--152137},
  year={2019},
}

@INPROCEEDINGS{10279198,
  author={Jouhari, Mohammed and Ibrahimi, Khalil and Othman, Jalel Ben and Amhoud, El Mehdi},
  booktitle={IEEE Inter. Conf. on Communications}, 
  title={{Deep Reinforcement Learning-Based Energy Efficiency Optimization for Flying LoRa Gateways}}, 
  year={2023}
}

@inproceedings{foerster2018counterfactual,
  title={Counterfactual multi-agent policy gradients},
  author={{Foerster, Jakob and Farquhar, Gregory and Afouras, Triantafyllos and Nardelli, Nantas and Whiteson, Shimon}},
  booktitle={Proceedings of the AAAI conference on artificial intelligence},
  volume={32},
  number={1},
  year={2018}
}

@article{sunehag2017value,
  title={{Value-decomposition networks for cooperative multi-agent learning}},
  author={Sunehag, Peter and others},
  journal={arXiv preprint arXiv:1706.05296},
  year={2017}
}

@article{rashid2020monotonic,
  title={{Monotonic value function factorisation for deep multi-agent reinforcement learning}},
  author={Rashid, Tabish and Samvelyan, Mikayel and De Witt, Christian Schroeder and Farquhar, Gregory and Foerster, Jakob and Whiteson, Shimon},
  journal={Journal of Machine Learning Research},
  volume={21},
  number={178},
  pages={1--51},
  year={2020}
}

@ARTICLE{9272315,
  author={Yu, Yi and Mroueh, Lina and Duchemin, Diane and Goursaud, Claire and Vivier, Guillaume and Gorce, Jean-Marie and Terré, Michel},
  journal={IEEE Access}, 
  title={{Adaptive Multi-Channels Allocation in LoRa Networks}}, 
  year={2020},
  volume={8},
  number={},
  pages={214177-214189},
  }

@INPROCEEDINGS{zhang2025novel,
  author={Zhang, Xiaodan and Fu, Xiaoying and Miao, Jiansong and Yao, Yushun and Mu, Junsheng},
  booktitle={2025 IEEE 101st Vehicular Technology Conference (VTC2025-Spring)}, 
  title={{A Novel Multi-Agent RL Approach in Priority-Aware UAV-Assisted Networks for AoI and Energy Consumption Minimization}}, 
  year={2025},
  volume={},
  number={},
  pages={1-6},
  }

@article{gong2023modeling,
  title={{Modeling power consumptions for multirotor UAVs}},
  author={Gong, Hao and Huang, Baoqi and Jia, Bing and Dai, Hansu},
  journal={IEEE Transactions on Aerospace and Electronic Systems},
  volume={59},
  number={6},
  pages={7409--7422},
  year={2023},
  publisher={IEEE}
}

@article{li2024collaborative,
  title={{Collaborative Task Offloading and Resource Allocation in Small-Cell MEC: A Multi-Agent PPO-Based Scheme}},
 author={Li, Han and others},
  journal={IEEE Trans. on Mobile Computing},
  year={2024},
  publisher={IEEE}
}

@ARTICLE{bazzi2502upper,
  author={Bazzi, Ahmad and others},
  journal={IEEE Communications Magazine}, 
  title={{Upper Mid-Band Spectrum for 6G: Vision, Opportunity and Challenges}}, 
  year={2026},
  volume={64},
  number={1},
  pages={206-212}}

@inproceedings{saliah2024multi,
  title={{Multi-sided matching for space-air-ground integrated systems}},
  author={Saliah, Abdoul Karim AH and Hamza, Doha and El Hammouti, Hajar and Shamma, Jeff S and Alouini, Mohamed-Slim},
  booktitle={2024 IEEE 99th Vehicular Technology Conference (VTC2024-Spring)},
  pages={1--7},
  year={2024},
  organization={IEEE}
}

@book{albrecht2024multi,
  title={{Multi-agent reinforcement learning: Foundations and modern approaches}},
  author={Albrecht, Stefano V and Christianos, Filippos and Sch{\"a}fer, Lukas},
  year={2024},
  publisher={MIT Press}
}
\end{document}